\documentclass[twocolumn,amsmath,amssymb,pra,superscriptaddress]{revtex4-1}
\usepackage[dvipdfmx]{graphicx}
\usepackage{dcolumn}
\usepackage{bm}
\usepackage{subfigure}
\usepackage{booktabs}
\usepackage{color}
\newcounter{one}
\def\ket#1{\mbox{\boldmath $#1$}}

\newcommand{\affA}{
Department of Mathematical and Computing Science, Tokyo Institute of Technology, 
W8-45, 2-12-1 Ookayma, Meguro-ku, Tokyo, Japan
}
\newcommand{\affB}{
Artificial Intelligence Research Center, 
National Institute of Advanced Industrial Science and Technology, 
2-3-26 Aomi, Koto-ku, Tokyo, Japan
}

\begin{document}

\title{\textbf{Evaluating network partitions through visualization}}

\author{Chihiro Noguchi}
\affiliation{\affA}
\author{Tatsuro Kawamoto}
\affiliation{\affB}
\date{\today}

\begin{abstract}
Network clustering requires making many decisions manually, such as the number of groups and a statistical model to be used. 
Even after filtering using an information criterion or regularizing with a nonparametric framework, we are commonly left with multiple candidates with reasonable partitions. 
In the end, the user has to decide which inferred groups should be regarded as informative. 
Here we propose a visualization method that efficiently represents network partitioning based on statistical inference algorithms. 
Our non-statistical assessment procedure based on visualization helps users extract informative groups when they cannot uniquely determine significant groups on the basis of statistical assessments. 
The proposed visualization is also effective for use as a benchmark test of different clustering algorithms. 
\end{abstract}
 
\maketitle

\section{\label{sec:Introduction}Introduction}
Network clustering is used to identify sets of structurally ``similar'' subgraphs. These are termed groups, modules, clusters, or communities. 
This is generally a nontrivial task because vertices are only statistically similar, and the network may not be clearly represented as a patchwork of groups. 
Some algorithms \cite{Nowicki2001,Peixoto2017tutorial,NewmanReinert2016,latouche12,Decelle2011,Decelle2011a,Hayashi2016,NewmanClauset2016} present a probability distribution describing the group assignment for each vertex, while some explicitly permit a vertex to belong to multiple groups simultaneously \cite{MMsbm,BallKarrerNewman2011,PeixotoPRX2015}. 

In this paper, we propose a visualization method that efficiently represents group assignment probability distributions of vertices in a network. 
The \textit{palette diagrams} that we introduce allow us to assess such high-dimensional information using manifold learning techniques \cite{LeeVerleysen2007}. 
An intuition behind is that, because a network clustering is essentially a dimensional reduction method, the corresponding probability distributions are expected
be nicely embedded in a low-dimensional manifold. 

We restrict our interest to network clustering methods such that each vertex may not strictly belong to a single group (i.e. soft clustering). 
[In this paper, we denote the resulting set of groups as a \textit{partition}, even when those groups overlap.] 
To draw a palette diagram, we take a set of partitions as the input, which may be partitions from different models and algorithms. 
The inputs can be different partitions from a single algorithm, e.g. partitions with different hyperparameters or partitions of different samples obtained with a Markov chain Monte Carlo (MCMC) algorithm. 


In the Results section, we introduce palette diagrams and explain how they can be used in practice. 
In the Methods section, we discuss a routine for drawing palette diagrams. 
The Discussion section is devoted to explaining how the palette diagrams offer a novel perspective for assessing network clustering.

\section{\label{sec:Results}Results}

\begin{figure*}[t!]
 \begin{center}
   \includegraphics[width= 2 \columnwidth]{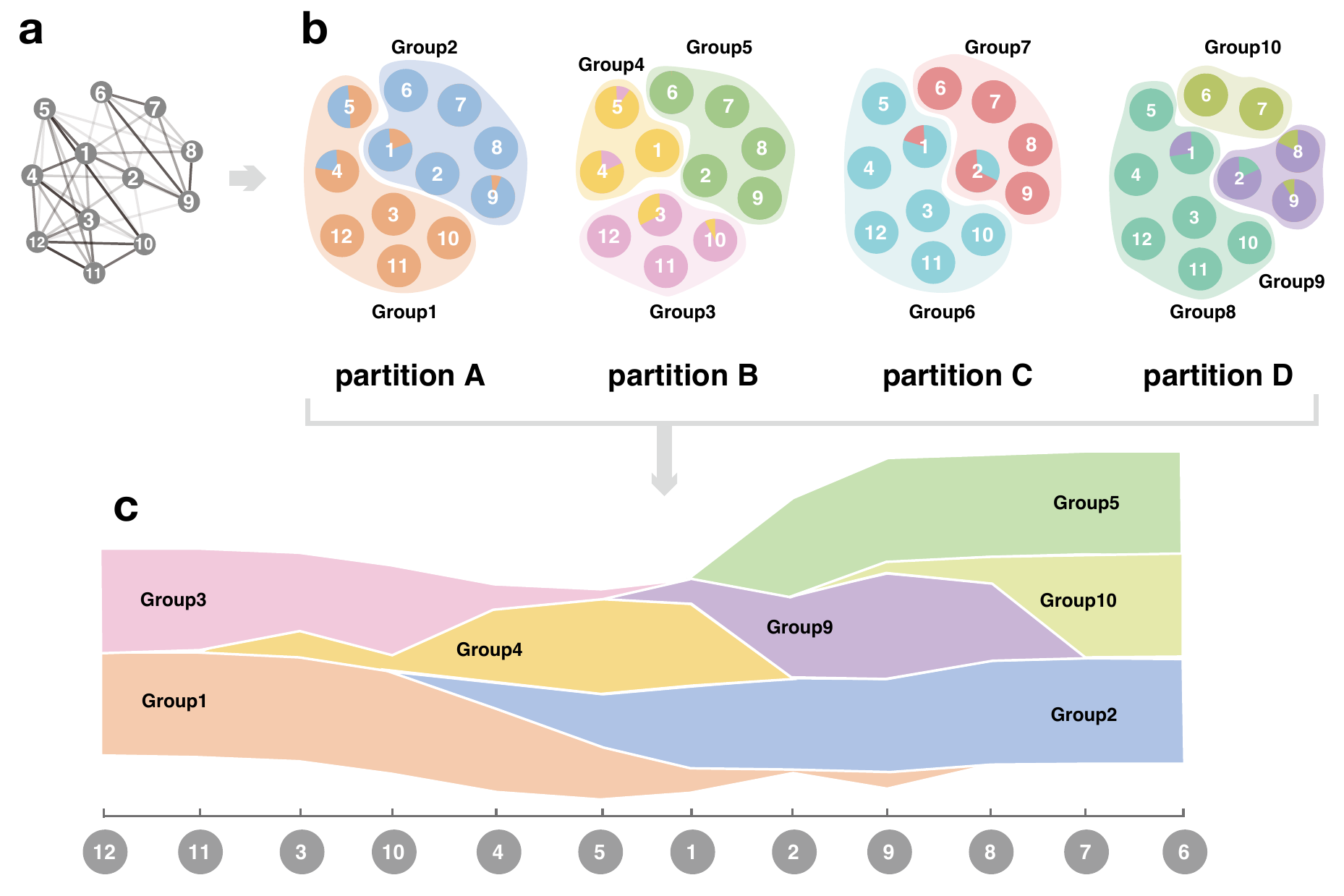}
 \end{center}
 \caption{
	\textbf{Schematic showing the construction of palette diagrams.} 
	(a) For a given network, (b) various partitions are obtained from network clustering algorithms based on statistical inference. 
	These results are high dimensional because they are obtained as a set of probability distributions for the vertices. 
	The group assignment probabilities are represented using the pie charts. 
	(c) In the (one-dimensional) palette diagram, the group assignment distribution for each vertex is minimally represented and the vertex ordering is optimized such that the modular structure can be visually interpreted. 
	}
 \label{fig:PaletteExample}
\end{figure*}

\subsection{\label{sec:Palette}Palette diagram}

Let us consider the network shown in Fig.~\ref{fig:PaletteExample}a.  
Figure \ref{fig:PaletteExample}b) shows multiple partitions (partition A -- partition D) corresponding to each clustering trial (e.g. with different values of hyperparameters). 
While these are often represented using pie charts, this visualization technique becomes increasingly impractical as the number of vertices increases and partitions overlap in a more complex manner. 
In addition, visualizing all groups is redundant because some of the identified groups are very similar (e.g. groups 1, 6, and 8, as well as groups 5 and 7). 
Figure \ref{fig:PaletteExample}c shows a (one-dimensional) palette diagram that will be explained in detail below.  
This visualization overcomes the above problems. 

The one dimensional palette diagram is plotted based on the streamgraph \cite{byron2008streamgraph}. 
The streamgraph usually shows how multidimensional data evolve over time. 
Whereas the horizontal axis represents timestamps in the streamgraph, the vertex labels are aligned horizontally in the palette diagram. 
The vertical degrees of freedom represent the group assignments. 
Each colour represents a group label, and its vertical width represents the magnitude of the assignment probability to that group. 

In the palette diagram, groups that represent qualitatively identical structures are filtered out. 
In Fig.~\ref{fig:PaletteExample}c, groups 6 and 8 are filtered out because they are similar to group 1. 
Furthermore, group 7 is filtered out because it is nearly identical to group 5. 
The vertical width of the diagram remains constant if all groups are kept in the diagram, indicating that the group assignment probabilities are normalized (e.g. see Figs.~\ref{fig:ManipulationComparison}a and \ref{fig:ManipulationComparison}b). 
In addition, as was the case in the streamgraph, the origin of the vertical axis is shifted for each vertex label. 

The two-dimensional palette diagram carries the same information as its one-dimensional counterpart. 
Instead of piling up the assignment probabilities of all groups vertically, each row shows the assignment probability of a group, and its magnitude is represented by the colour depth. 
While the two-dimensional palette diagram has an advantage in that similarities among the group assignment patterns can be inspected more easily, the one-dimensional counterpart has an advantage in that the boundaries of overlapping groups can be inspected more easily.

\subsection{Effects of vertex ordering and group filtering}

\begin{figure*}[t!]
 \begin{center}
   \includegraphics[width= 1.6 \columnwidth]{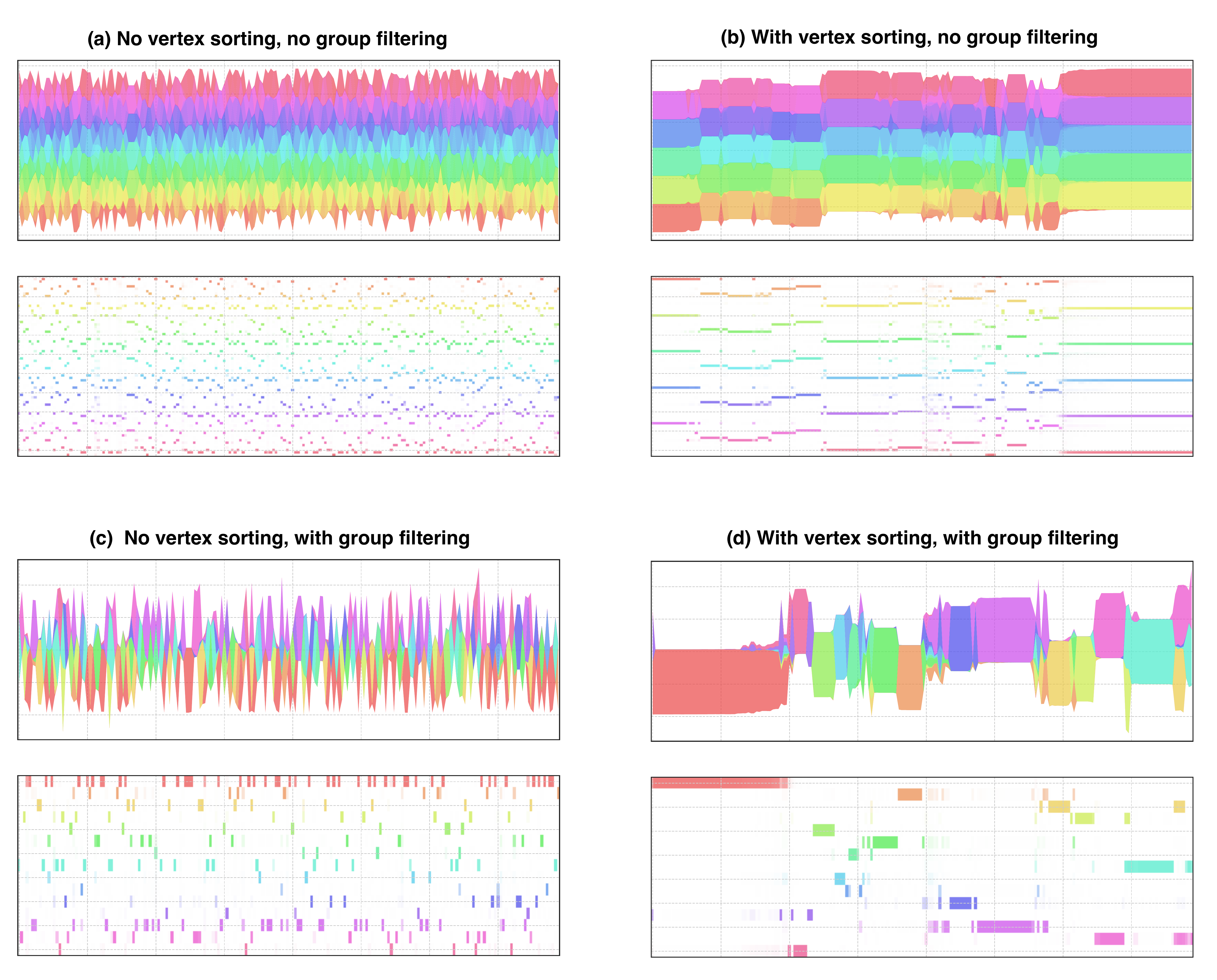}
 \end{center}
 \caption{
	\textbf{Palette diagrams for the \textit{jazz} network with/without group filtering and vertex sorting.} 
	}
 \label{fig:ManipulationComparison}
\end{figure*}

There is no intrinsic ordering for vertices unlike the usual data considered in a streamgraph. 
Aside from synthetic networks or networks that are carefully sorted based on metadata, vertices of successive labels do not necessarily tend to belong to the same group. 
Therefore, the vertices must be optimally aligned so that the group assignment probabilities vary smoothly. 

We demonstrate the effect of two important manipulations in drawing palette diagrams: group filtering and vertex sorting. 
We consider a collaboration network (\textit{jazz} network) between jazz musicians in which the vertices represent jazz musicians and the edges represent the level of collaboration \cite{konect:2016:arenas-jazz,konect:arenas-jazz}. 
We use a nonparametric MCMC proposed in Ref.~\cite{NewmanReinert2016} as a clustering algorithm. 

When the palette diagrams are not manipulated (Fig.~\ref{fig:ManipulationComparison}a), the palette diagrams look very noisy and we can hardly identify the structures therein. 
When the order of the vertices is optimized (Fig.~\ref{fig:ManipulationComparison}b), we can confirm that the algorithm identifies a modular structure. 
However, while the groups with similar colour codes represent essentially different groups, the set of groups with different colours represent essentially identical partitions. 
Thus, the diagram is very redundant. 
Let us then consider the case where the redundant groups are filtered out without optimized vertex ordering (Fig.~\ref{fig:ManipulationComparison}c). 
Different colours are now assigned to essentially different groups, but again, it is difficult to identify the modular structure. 
Finally, the modular structure is minimally represented in an interpretable manner when both vertex sorting and vertex ordering are performed (Fig.~\ref{fig:ManipulationComparison}d). 

Note that, while vertices may have very complex group assignments, the order of the vertices is the only degree of freedom that we arrange. 
Therefore, in principle, sometimes we cannot localize every group in a single region while keeping the boundaries of the groups smooth. 

The next section discusses how to distinguish between redundant and qualitatively different groups. 
The filtering resolution is parametrized as the number of groups to be focused on.

\subsection{\label{sec:GroupNumberAssessment}Assessment of partitions with different numbers of groups}

\begin{figure*}[t!]
 \begin{center}
   \includegraphics[width= 1.6\columnwidth]{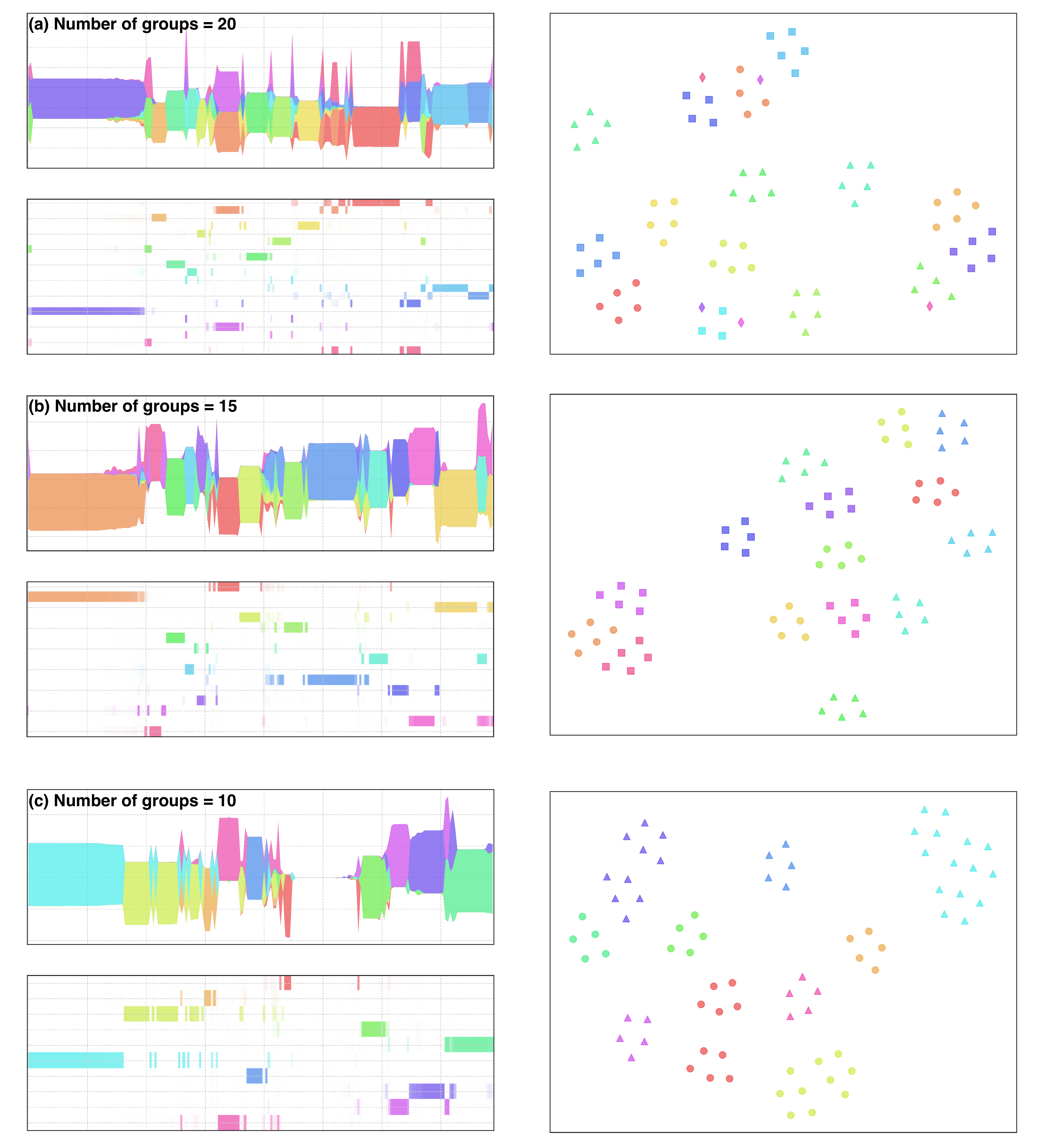}
 \end{center}
 \caption{
	\textbf{Palette diagrams for the \textit{jazz} network with different number of groups.} 
	One-dimensional and two-dimensional palette diagrams are presented on the left-hand side. 
	t-SNE plots for all identified groups are presented on the right-hand side. 
	Although the way in which the points are clustered is different and depends on the number of the groups to be selected, the distributions of these t-SNE plots are identical. 
	}
 \label{fig:GroupNumber}
\end{figure*}



The number of groups is specified as an input in some algorithms. 
Others estimate the number of groups and attempt to infer the corresponding partition. 
Within the framework of statistical modelling, this is a part of the model selection problem. 
The optimal number of groups is typically determined using a scalar criterion, such as modularity \cite{NewmanGirvan2004}, map equation \cite{Rosvall2008,Rosvall2011}, Bethe free energy \cite{Decelle2011a}, ICL and its variants \cite{latouche12,Hayashi2016}, minimum description length \cite{Peixoto2013,PeixotoPRX2014}, spectral band \cite{Luxburg2007,Krzakala2013,modBIX}, and cross-validation error \cite{KawamotosbmBIX, modBIX}, to name a few. 
For a given criterion, the optimal number of groups is chosen as the one that yields the extremum value of the criterion, or the one that is most parsimonious. 
Alternatively, nonparametric methods \cite{NewmanReinert2016,Peixoto2017PRE} based on MCMCs yields the plausible numbers of groups through sampling. 
Some of these methods are based on different principles, heuristics, or prior distributions. 

It should be emphasized that, in practice, one typically obtains multiple candidates for the most significant partition instead of a single optimum partition. 
It is not obvious which of these candidate partitions should be regarded as the most parsimonious one under a given criterion, and they are frequently sampled while executing nonparametric MCMCs. 
These partitions may be of qualitatively different structures, qualitatively identical, or may only have different resolution within a hierarchical modular structure. 
Usually, we would not realize these facts unless we directly evaluate the way the network is actually partitioned. 
Thus, although the assessment using a scalar criterion is important to narrow down the candidate partitions, visualizing the actual partitions as a final distillation process greatly aids in determining the groups that should receive focus.

Figure \ref{fig:GroupNumber} shows palette diagrams with different number of groups. 
The t-SNE plot for all identified groups is presented on the right in each figure; each point in the plot represents an identified group and the distance between a pair of groups is measured based on the Kullback--Leibler divergence of their probability distributions. 
We can see that some groups form a cluster in a t-SNE plot, indicating that these groups are similar. 
Therefore, we choose some representative groups and include them in the palette diagram. 
The details of the manipulation here are explained in Sec.~\ref{sec:filtering}.

In Fig.~\ref{fig:GroupNumber}, from top to bottom, the number of groups are set to 20, 15, and 10, respectively. 
The result with 15 groups exhibits an almost non-overlapping structure. 
When we increase the number of groups to 20, some groups appear to be redundant, while other groups are subdivided into smaller groups. 
When we decrease the number of groups to 10, some vertices no longer assigned to any group. 
Thus, we should use more than 10 groups. 

We cannot universally define the optimum number of groups. 
Redundancy is acceptable as long as a user can visually identify the modular structure. 
In the end, the number of groups a user can afford is the user's choice.

\subsection{\label{sec:ModelAlgorithmComparison}Comparison of different models and algorithms}

\begin{figure}[ht!]
 \begin{center}
   \includegraphics[width= 0.72 \columnwidth]{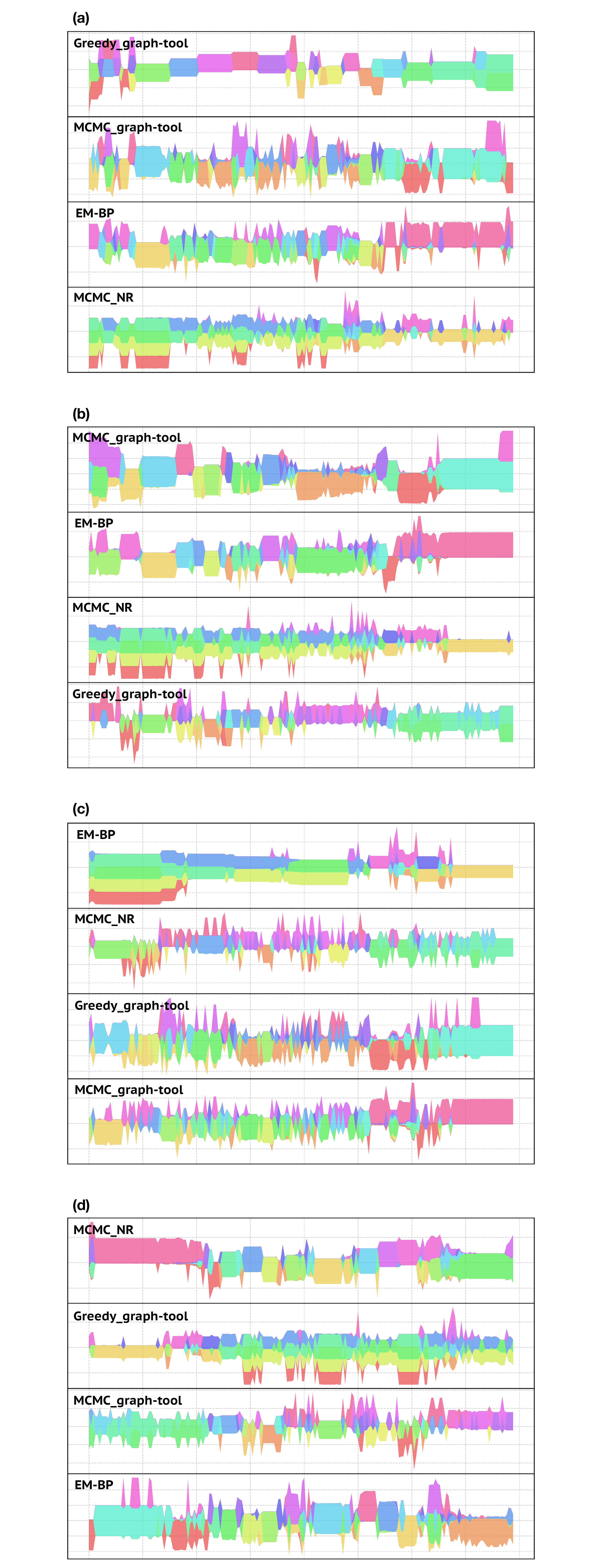}
 \end{center}
 \caption{
	\textbf{Comparison of palette diagrams generated from the \textit{jazz} network with the greedy, EM, and MCMC algorithms.}
	}
 \label{fig:jazzComparison}
\end{figure}

Finally, the palette diagrams are also effective for comparing the performance of different statistical models and algorithms. 
In fact, there are many related questions that were not investigated. 
How do the results from the expectation-maximization (EM) and MCMC algorithms differ for a given model or for models with slightly different formulation? 
How do the results from overlapping models compare with those from non-overlapping models? 
Practically, it is also important to confirm the performance of different implementations of the same algorithm. 

A comparison between a greedy algorithm, an EM algorithm, and two MCMCs is presented in Fig.~\ref{fig:jazzComparison}. 
Here, we refer to the EM algorithm considered in Ref.~\cite{Decelle2011a} as \textsf{EM-BP}, the MCMC considered in Ref.~\cite{NewmanReinert2016} as \textsf{MCMC\_NR}, and the MCMC considered in Ref.~\cite{PeixotoPRX2015} and implemented in graph-tool \cite{graphtool} as \textsf{MCMC\_graph-tool}. 
The greedy algorithm that we consider is also an implementation in graph-tool, which corresponds to the zero temperature limit of \textsf{MCMC\_graph-tool}; we refer to this algorithm as \textsf{Greedy\_graph-tool}. 
We used the code distributed at \cite{BIX-Github} to implement \textsf{EM-BP}. 
Among these algorithms, \textsf{Greedy\_graph-tool} and \textsf{MCMC\_graph-tool} employ an overlapping model while other two employ a non-overlapping model. 

The diagrams in Fig.~\ref{fig:jazzComparison}a represent four one-dimensional palette diagrams in which vertex ordering is optimized with respect to the result from \textsf{Greedy\_graph-tool}. 
Analogously, in Figs.~\ref{fig:jazzComparison}b, \ref{fig:jazzComparison}c, and \ref{fig:jazzComparison}d, the vertex ordering is optimized with respect to \textsf{MCMC\_graph-tool}, \textsf{EM-BP}, and \textsf{MCMC\_NR}, respectively. 

One can confirm that, for this particular dataset (the \textit{jazz} network) and the hyperparameter settings we chose, all results are somewhat consistent. 
The result from \textsf{EM-BP} has lower resolution compared to the others. 
While \textsf{EM-BP} and \textsf{MCMC\_graph-tool} exhibit hierarchical structures, \textsf{Greedy\_graph-tool} and \textsf{MCMC\_NR} exhibit non-hierarchical structures. 
A prominent overlapping structure is not observed regardless of the model we choose. 


Comparison of the palette diagrams from other real-world networks is shown at \cite{PaletteExamples}.


\section{\label{sec:Discussion}Discussion}
The palette diagrams provide visualizations of the group assignment probability distributions. 
While many statistical inference frameworks and algorithms have been proposed, an investigation into the output high-dimensional information is understudied. 

We showed that the palette diagrams are useful in many ways. 
These diagrams are useful for evaluating weak and overlapping modular structures, assessing significant groups, and comparing clustering results. 
The assessment of significant groups here relies on consistency and interpretability (i.e. non-statistical), and is distinct from that in terms of statistical significance. 
We also note that a visual assessment of a significant modular structure was also proposed in Ref.~\cite{modBIX}. 

A palette diagram offers a novel perspective for evaluating significant groups. 
We assess the significance of individual groups independently when we draw a palette diagram. 
Contrastingly, we typically evaluate the significance of a partition rather than the individual groups therein. 
In other words, the minimal element of the groups is constrained by the partitions. 
We refer to the former as a \textit{group-wise} assessment and the latter as a \textit{partition-wise} assessment. 
We would miss many informative groups that belong to multiple different partitions when performing a partition-wise assessment. 
This occurs prominently whenever a clustering algorithm exhibits a non-hierarchical structure. 
We note that group-wise assessment is different from detecting ``local communities'' \cite{Fortunato201075}; they are usually evaluated based only on local information in a network, without comparing with other inferred groups or referring to the total size of the network.  

We also note that the idea of comparing results from multiple algorithms is completely different from so-called consensus clustering \cite{lancichinetti2012consensus}. 
We compare results to determine whether the different algorithms produce consistent results. 
Each group is identified using a single algorithm based on a single principle: the groups themselves should not be modified by mixing multiple principles. 


The code for generating the palette diagrams is distributed at \cite{PaletteDiagramCode} and was implemented in Python.

\section{\label{sec:Methods}Methods}

\subsection{\label{sec:HowToDraw}Generating palette diagrams}
In this section, we explain how to generate one-dimensional and two-dimensional palette diagrams. 
They contain the same information, but each diagram displays this information differently. 


Suppose we have multiple partitions of the same network with $N$ vertices. 
We denote the number of groups in the $\ell$th partition as $M_{\ell}$ and the total number of groups of all the partitions as $M_{0}$, i.e. $M_{0} = \sum_{\ell=1}^{L} M_{\ell}$. 
Each individual partition can have a different number of groups. 
We combine these partitions to construct an $N \times M_{0}$ matrix $\ket{P_{0}}$. 
An element $p_{g,i}$ ($g \in \{1,2,\dots,M_{0}\}$, $i \in \{1,2,\dots,N\}$) in $\ket{P_{0}}$ represents the assignment probability that vertex $i$ belongs to group $g$. 
We denote this matrix as the assignment matrix and a row in $\ket{P}_{0}$, $\ket{p}_{g} := \{p_{g,1}, p_{g,2}, \dots, p_{g,N}\}$, as the assignment vector.



The raw assignment matrix $\ket{P_{0}}$ is not directly useful. 
As we mentioned in the Results section, $\ket{P_{0}}$ is redundant because some of the $M_{0}$ groups may essentially represent the same group. 
Moreover, vertices (i.e. columns) must be nicely ordered such that neighbouring vertices have similar assignment probabilities to facilitate a visual interpretation of the partitions. 
Therefore, our major task for visualization is to construct an $N \times M$ ($M \le M_{0}$) reduced assignment matrix $\ket{P}$ from $\ket{P_{0}}$. This is performed using 
(i) filtering of redundant groups and (ii) vertex sorting. 

\subsubsection{\label{sec:filtering}Filtering of redundant groups}
Among $M_{0}$ groups in the candidate partitions, we extract the minimal set of distinct groups ($M$ groups). 
To this end, we cluster the $M_{0}$ groups. 
We define the similarity between a pair of group assignment distributions using an $\alpha$-divergence \cite{Cichocki2009}. 
$\alpha$-divergences are general distance measures between distributions. 
By varying the values of $\alpha$, we can modify the measure of the total variation, L2 norm, Kullback--Leibler divergence, Shannon divergence, and a number of other measures. 

We specify a pair of values $(\alpha, M)$. 
We construct an $M_{0}\times M_{0}$ similarity matrix based on the $\alpha$-divergence. 
We then $M$ partition this similarity matrix using k-means clustering (i.e. clustering of the clustering results). 
Finally, to confirm whether our choice of $(\alpha, M)$ is plausible, we visualize the result from k-means clustering using the 2 dimensional t-SNE (e.g. see the right panels in Fig.~\ref{fig:GroupNumber}). 
If the result is implausible, we repeat the above process for different values of $(\alpha, M)$. 

After $M$ partitioning of $M_{0}$ groups, we extract the $M$ groups on which we focus. 
From each group in the $M$ partitioning (i.e. a group of groups), we select the one group such that $\sum_{i=1}^{N} p_{gi}$ is maximum as a representative group.


\subsubsection{\label{sec:sorting}Vertex sorting}
After we reduce the set of groups to receive focus, we sort the ordering of the vertices such that the partitions can be easily assessed. 
To this end, we used the Isomap \cite{ISOMAP_Tenenbaum,LeeVerleysen2007}, a nonlinear dimensional reduction method from manifold learning. 
We originally have $N$ data points in the $M$ dimensional space based on the reduced assignment matrix $\ket{P} = [p_{g, i}]$. The Isomap embeds these points into a one dimensional space, which yields the order of vertices that reflects group assignment similarities. 

As we briefly mentioned in the Results section, it is not always possible to align vertices perfectly such that group assignments for each group are gathered in a single region. 
We might need to place another group in two separated regions to force one group assignment gathered in a single region. 
This reflects the fact that, in some cases, a modular structure essentially exists in a higher dimensional manifold. 
This trade-off between good interpretability and correct representation is commonly encountered in manifold learning \cite{LeeVerleysen2007}.  
Optimizing this sorting becomes more difficult as the number of groups $M$ becomes larger and the partitions represent a more complex structure (i.e. non-hierarchical and overlapping structures). 
The choice of algorithm for vertex sorting is critical for visual assessment and, as far as we investigated, the Isomap is sufficient for the present problem. 


\subsection{\label{sec:PaletteDiagramConstruction}Constructing one-dimensional and two-dimensional palette diagrams}
The above steps yield the reduced assignment matrix $\ket{P}$ with optimized ordering. 
The two dimensional palette diagram simply represents this matrix in which different colours are assigned for each group. 
In the two dimensional palette diagram, the assignment probability for each group is represented by the colour depth, while it is represented by the height along the vertical direction in the one-dimensional palette diagram. 


\section*{Acknowledgments}
This work was supported by JSPS KAKENHI No. 17H06103.

\bibliographystyle{apsrev}
\bibliography{bib-palette}
\end{document}